\documentclass[11pt]{article}
\usepackage[T1]{fontenc}
\usepackage{lmodern}
\usepackage[margin=1in]{geometry}
\usepackage{amsmath,amssymb,amsthm}
\usepackage{microtype}
\usepackage{enumitem}
\usepackage{float}
\usepackage{tabularx,booktabs}
\usepackage{needspace}
\usepackage[numbers,sort&compress]{natbib}
\usepackage[hidelinks]{hyperref}
\usepackage{xcolor}
\usepackage{authblk}

\newif\ifshowchanges
\showchangesfalse
\newcommand{\editcolor}{\ifshowchanges\color{blue}\else\color{black}\fi}
\DeclareRobustCommand{\edit}[1]{{\editcolor #1}}
\newenvironment{revision}{\par\begingroup\editcolor}{\par\endgroup}
\hypersetup{pdftitle={Two (2+epsilon)-Approximation Algorithms for Minimum Spanning Tree Interdiction},pdfsubject={Approximation algorithms for minimum spanning tree interdiction},pdfkeywords={MST interdiction, approximation algorithms, graph cuts, small set expansion}}
\newtheorem{theorem}{Theorem}[section]
\newtheorem{lemma}[theorem]{Lemma}
\newtheorem{corollary}[theorem]{Corollary}
\newtheorem{proposition}[theorem]{Proposition}

\theoremstyle{remark}

\DeclareMathOperator{\MST}{MST}
\DeclareMathOperator{\OPT}{OPT}
\DeclareMathOperator{\TSP}{TSP}
\floatstyle{ruled}
\newfloat{algorithm}{tbp}{loa}
\floatname{algorithm}{Algorithm}
\setlist[enumerate]{leftmargin=*,itemsep=3pt,topsep=4pt}
\setlist[itemize]{leftmargin=*,itemsep=3pt,topsep=4pt}
\title{\edit{Two $(2+\epsilon)$-Approximation Algorithms for
Minimum Spanning Tree Interdiction}}
\author[1]{Junjie Luo}
\author[2]{Changjun Wang}
\author[3]{Chenhao Wang}
\author[4]{Wei Yu}

\affil[1]{School of Mathematics and Statistics, Beijing Jiaotong University, Beijing, China\\
  \texttt{jjluo1@bjtu.edu.cn}}

\affil[2]{SKLMS, Academy of Mathematics and Systems Science, Chinese Academy of Sciences, Beijing, China\\
  \texttt{wcj@amss.ac.cn}}

\affil[3]{Beijing Normal University-Zhuhai, Zhuhai, China\\
  \texttt{chenhwang@bnu.edu.cn}}

\affil[4]{School of Mathematics, East China University of Science and Technology, Shanghai, China\\
  \texttt{yuwei@ecust.edu.cn}}
\date{}
\begin{document}
\maketitle
\vspace{-2em}
\begin{abstract}
\edit{Minimum spanning tree interdiction asks for edge deletions within a
budget that maximize the weight of a remaining minimum spanning tree.
We give two deterministic $(2+\epsilon)$-approximation algorithms for
every fixed $\epsilon>0$, improving the previous best factor of $4$.
Both algorithms respect the original budget and run in
$N^{O(1/\epsilon)}$ time for $0<\epsilon\le1$, where $N$ is the binary
input size. The first algorithm uses two weight thresholds to choose
initial deletions that force every remaining spanning tree to contain
edges at or above these thresholds. It then uses the remaining budget
to further increase the MST weight through Lagrangian rounding.
The second algorithm achieves the same guarantee through a different
decomposition: it enumerates vertex partitions and heavy trees
connecting their parts, then approximates a bounded-weight interdiction
problem within the parts. An objective-preserving reduction from
Maximum Components shows that, assuming the Small Set Expansion
Hypothesis and $\mathsf P\ne\mathsf{NP}$, no deterministic
polynomial-time $(2-\delta)$-approximation exists for any fixed
$0<\delta<1$, even on simple graphs with weights in
$\{0,1\}$. Thus, our approximation guarantee is conditionally tight up to 
an arbitrarily small additive constant. Finally, the MST guarantee yields a
$(4+\epsilon)$-approximation for metric-TSP interdiction.}
\end{abstract}

\section{Introduction}
\label{sec:introduction}

\begin{revision}
A network can remain connected after several links fail and yet become
substantially more expensive to operate. When the operator must connect
all vertices, a minimum spanning tree (MST) measures the cheapest surviving
connection. Minimum spanning tree interdiction asks which links an
adversary should remove, subject to a deletion budget, to make this
connection as expensive as possible. It thus measures vulnerability
beyond whether the network can be disconnected.

Formally, let $G=(V,E)$ be a connected undirected graph with $n$ vertices. Each edge $e\in E$
has a nonnegative weight $w_e$ and a nonnegative deletion cost $c_e$.
The weight contributes to the operator's objective, while the deletion
cost contributes to the adversary's budget. These quantities need not
be related: an edge that is inexpensive to use may be expensive to remove.
Given a budget $B$, the task is to choose $R\subseteq E$ with
$c(R):=\sum_{e\in R}c_e\le B$ maximizing
$F(R):=\MST_w(G-R)$, the remaining MST weight. A minimum-cut computation
detects whether an affordable deletion disconnects $G$. We therefore
focus on instances in which every feasible deletion leaves the graph
connected. Write $\OPT=\max_{c(R)\le B}F(R)$. An $\alpha$-approximation
returns a feasible deletion $R$ with $F(R)\ge\OPT/\alpha$.

MST interdiction is a classical and well-studied problem in network
interdiction, with early work dating back to Lin and Chern~\cite{LC93} and Frederickson and Solis-Oba~\cite{FS99}.
Although computing an MST is polynomial-time solvable, finding an
optimal interdiction is NP-hard. 
Early logarithmic approximation guarantees~\cite{FS99} were improved
to a factor of $14$ by Zenklusen~\cite{Zen15} and subsequently to $4$ 
by Linhares and Swamy~\cite{LS17}.
\end{revision}

\subsection{Our results}

\edit{We give two deterministic algorithms that improve the approximation
ratio from $4$ to $2+\epsilon$ without exceeding the deletion budget.
Both algorithms achieve the following guarantee.}

\begin{theorem}
\label{thm:approximation}
For every fixed $\epsilon>0$, MST interdiction admits a deterministic
$(2+\epsilon)$-approximation that respects the deletion budget.
For $0<\epsilon\le1$, its running time is $N^{O(1/\epsilon)}$, where
$N$ is the binary input size.
\end{theorem}

\edit{The first algorithm uses two weight thresholds to create enough
MST value before rounding. The second separates heavy connections
between parts from optimization within the parts. We develop both
algorithms in the main text, in Sections~\ref{sec:approximation}
and~\ref{sec:quotient}, respectively.}

\begin{revision}
This factor approaches a conditional barrier. The Maximum Components
hardness of Aissi, Attias, Chen, and Ravi~\cite[Theorem~5]{AACR22}
transfers to MST interdiction through a construction that preserves
objective values exactly. Under the Small Set Expansion Hypothesis
(SSEH), this excludes a deterministic polynomial-time
$(2-\delta)$-approximation for every fixed $0<\delta<1$, unless
$\mathsf P=\mathsf{NP}$. The result holds even on simple graphs with
weights in $\{0,1\}$ that remain connected after every feasible deletion
(Theorem~\ref{thm:hardness}).
\end{revision}

Finally, the relation between spanning trees and covering closed walks
gives a $(4+\epsilon)$-approximation for metric-TSP interdiction
(Corollary~\ref{cor:tsp}). Here the operator finds a cheapest closed walk
visiting every vertex after the deletion, equivalently a TSP tour in
the shortest-path metric of the remaining graph.

\begin{revision}
\subsection{Algorithm overview and main ideas}
\label{subsec:technical-contributions}
Both algorithms combine cut enumeration, based on the framework
of~\cite{CQX20}, with the Lagrangian rounding framework of Linhares and Swamy~\cite{LS17}. Section~\ref{sec:enumeration} states the cut-enumeration tools used below.
The algorithms control the additive rounding loss in different
ways: the two-threshold algorithm creates enough MST value
before rounding, while the heavy quotient tree algorithm
compares the rounded solution with a separate threshold-cut
candidate.

\paragraph{The two-threshold algorithm.}
For each pair $a\ge t>0$ of input edge weights, the algorithm generates
candidate initial deletions $C$ that force every remaining spanning
tree to contain at least two edges of weight at least $t$, including
one of weight at least $a$. For each $C$, it caps the surviving edge
weights at $t$ and uses the remaining budget to extend $C$ through
Lagrangian rounding. It evaluates the resulting deletions under the
original edge weights and returns the best candidate.

Linhares and Swamy~\cite{LS17} use a hierarchy of components to round
two nested Lagrangian solutions. Their tree-knapsack procedure keeps
the smaller deletion and extends it using the remaining budget.
On the original instance, their Lagrangian upper bound can be nearly
three times the optimum~\cite[full version, Theorem~4.10]{LS17}.
Thus rounding directly against that bound does not explain a factor
near $2$. We apply the rounding framework after a carefully chosen
initial deletion.

For the residual instance after deleting $C$, let $D=B-c(C)$,
let $L(D)$ be the residual Lagrangian upper bound, and write $h(S)$
for the MST value in the auxiliary graph after a further deletion $S$.
Lemma~\ref{lem:baseline} gives
\[
  h(S)\ge\frac{L(D)+h(\varnothing)}2-t.
\]
Thus, if the initial deletion guarantees $h(\varnothing)\ge2t$,
its contribution covers the rounding loss $t$.

Fix an optimal deletion $R^\star$. If its remaining MST has at least
two positive-weight edges, the algorithm tries its two largest edge
weights $a\ge t$, counted with multiplicity. The initial deletion $C$
leaves at least three components in the subgraph of edges lighter than
$t$ and at least two in the subgraph of edges lighter than $a$.
The first condition gives an initial capped MST value
$h(\varnothing)\ge2t$. Only the largest edge of the optimal comparison
tree can exceed $t$, so capping this tree loses exactly $a-t$.
The second condition ensures that every spanning tree remaining after
rounding still contains an edge of original weight at least $a$.
Restoring the original weights therefore recovers at least $a-t$.

An affordable initial deletion can still leave too little budget for
rounding. To control this cost, the algorithm constructs each $C$ as
the union of two disjoint sets $P$ and $Z$. For the threshold pair
just described and a parameter $0<\eta<1$, Lemma~\ref{lem:completion}
guarantees at least one candidate with
\[
  P\subseteq R^\star,\qquad c(Z)\le\eta\bigl(B-c(P)\bigr).
\]
The algorithm either completes a candidate with low-cost cuts or
branches on a near-minimum cut~\cite{CQX20}. Each branching step
settles one of the two threshold conditions.

Only $Z$ can cost extra relative to $R^\star$. The remaining eligible
edges of $R^\star$ may exceed the residual budget $D$, but the
Lagrangian relaxation still includes this set. Comparing it with the
empty deletion, rounding, and restoring the original weights gives
a value of at least $(1-\eta)\OPT/2$. Setting
$\eta=\epsilon/(2+\epsilon)$ gives the claimed ratio.

\paragraph{The heavy quotient tree algorithm.}
Fix an integer $k\ge1$ and suppose $n\ge k+1$.
The algorithm chooses a threshold $q$ with two properties:
every feasible deletion leaves at most $k$ components in the
subgraph of edges of weight at most $q$, and a feasible
threshold-cut candidate has MST value at least $kq$.
The component bound allows the algorithm to enumerate all
affordable vertex partitions of this subgraph and all trees
of edges heavier than $q$ connecting their parts.

For each partition and heavy tree, the algorithm deletes
inter-part edges that would prevent the chosen tree from
being a minimum spanning tree of the quotient graph.
It discards choices whose deletion cost exceeds the budget.
For each remaining choice, it uses the remaining budget to
optimize deletions within the parts through Lagrangian
rounding on a bounded-weight auxiliary instance.

The key comparison uses a branch corresponding to an optimal
deletion $R^\star$. On this branch, the initial deletion is
contained in $R^\star$, so the remaining optimal internal
deletion still fits the residual budget.
The auxiliary instance preserves the internal contribution
of the optimum, while the heavy tree accounts for the
connections between parts.
This gives a candidate of value at least $\OPT/2-q$.

Let $\mathrm{ALG}$ be the largest value among all candidates.
The rounded candidate and the threshold-cut candidate give
\[
  \mathrm{ALG}\ge\frac{\OPT}{2}-q,
  \qquad
  \mathrm{ALG}\ge kq.
\]
Thus $\OPT\le(2+2/k)\mathrm{ALG}$.
Choosing $k=\lceil2/\epsilon\rceil$ gives the claimed
$(2+\epsilon)$-approximation; instances with $n\le k$
are solved exactly.

\end{revision}

\subsection{Related work}


\begin{revision}
With unit deletion costs, MST interdiction is known as the $k$ most vital edges problem, studied by Frederickson and Solis-Oba~\cite{FS99}. The special case with two distinct edge weights captures budgeted graph disconnection, for which Engelberg et al.~\cite{EKLN07} established a $2$-approximation. More generally, MST interdiction is closely related to Maximum Components and can be viewed as a multi-level graph disconnection problem~\cite{Zen15,LS17}.
\end{revision}

Other spanning-tree interdiction objectives require different
approximation guarantees. Ostrovsky, Rabani, and Siman Tov~\cite{ORT25}
study budget minimization and maximization of the increase in MST weight.
Subtracting the original MST value preserves the optimal deletion sets
but changes multiplicative guarantees. Exact solution methods include
the integer programming formulations for the minimum-cost blocker
variant by Wei, Walteros, and Mahdavi Pajouh~\cite{WWP21} and the MST
and matroid basis interdiction algorithms of Weninger and
Fukasawa~\cite{WF26}.

Goldschmidt and Hochbaum~\cite{GH94} established polynomial-time
solvability of minimum $k$-cut for fixed $k$. Chekuri, Quanrud, and
Xu~\cite{CQX20} developed an LP and tree-packing framework for cut
algorithms and near-minimum-cut enumeration. 
Building on this framework, we derive an enumeration method for 
affordable vertex partitions, which is used by both of our algorithms. 
Our residual rounding procedure relies on the Lagrangian framework 
of Linhares and Swamy~\cite{LS17}, together with submodular function
minimization~\cite{IFF01,Sch00} and the equivalence
between separation and optimization~\cite{GLS81}.

For hardness, Raghavendra and Steurer~\cite{RS10} introduced SSEH,
and Raghavendra, Steurer, and Tulsiani~\cite{RST12} developed
reductions between expansion problems.
Manurangsi~\cite{Man18} established SSEH-based hardness for minimum
$k$-cut using a small-parts counting argument.
Aissi, Attias, Chen, and Ravi~\cite{AACR22} established SSEH-based 
hardness for Maximum Components,
which provides the starting point for our objective-preserving reduction.
The metric-TSP consequence uses the spanning-tree and covering-walk
comparison of earlier interdiction algorithms~\cite{Zen15,LS17}.

\begin{revision}
\paragraph{Organization.}
Section~\ref{sec:preliminaries} gives the model, threshold identities,
cut-enumeration tools, and residual rounding bound.
Section~\ref{sec:approximation} presents the two-threshold algorithm,
explains how to generate its initial deletions, and proves its
approximation guarantee and rounding bound.
Section~\ref{sec:quotient} gives the second algorithm, including its
exact routine for small instances.
Section~\ref{sec:consequences} treats conditional hardness and
metric-TSP interdiction. The appendices contain the forest-packing
proof, the technical tools for residual rounding, and the full
hardness proof.
\end{revision}

\section{Preliminaries}
\label{sec:preliminaries}

Let $G=(V,E)$ be a connected undirected graph with $n$ vertices and $m$
edges. Parallel edges are allowed, and loops can be discarded. Each
edge $e\in E$ has a nonnegative weight $w_e$ and a nonnegative deletion
cost $c_e$. All numerical input data are rational and encoded in
binary; let $N$ denote the total encoding length.
For any $R\subseteq E$, let $G-R=(V,E\setminus R)$ be the
graph obtained by deleting the edges in $R$, and write
$c(R)=\sum_{e\in R}c_e$.

Let $\MST_w(H)$ denote the weight of a minimum spanning tree of $H$,
with value $+\infty$ if $H$ is disconnected and value zero if $H$ has
a single vertex. Given an interdiction budget $B\ge0$, 
a deletion
$R\subseteq E$ is feasible if $c(R)\le B$. Define
\[
F(R)=\MST_w(G-R),\qquad
\OPT=\max_{R\subseteq E:\,c(R)\le B}F(R).
\]
The objective of MST interdiction is to find a feasible deletion
maximizing $F(R)$.

For $\alpha\ge1$, an $\alpha$-approximation algorithm returns a feasible
deletion $R$ satisfying $F(R)\ge\OPT/\alpha$ whenever $\OPT$ is finite.
If $\OPT=+\infty$, it returns a feasible deletion that disconnects $G$.

The case $n=1$ is trivial. Delete all zero-cost edges and include them
in the final output; this cannot decrease the MST value. Return these
deletions if the graph becomes disconnected or $B=0$. Otherwise,
a minimum cut with edge costs $c_e$ detects whether an affordable
disconnection exists, which we return together with the zero-cost
deletions. Henceforth, we assume $n\ge2$, $B>0$, positive deletion
costs, and that every feasible deletion leaves the graph connected.

For $t\ge0$, let $G_{<t}$ retain all vertices of $G$ and only
edges with $w_e<t$; define $G_{\le t}$ similarly using $w_e\le t$.
Write $\kappa(H)$ for the number of connected components of a graph
$H$, including isolated vertices.

We also use a separate operation called \emph{capping}. After deleting an
edge set $R$, reduce any weight of remaining edges greater than $t$ to $t$, such that
 each edge $e$ receives weight $\min\{w_e,t\}$.
Let $F_t(R)$ denote the minimum spanning tree weight of $G-R$ computed with these modified weights. Thus threshold graphs select edges, whereas capping changes their weights.

If edges of weight less than \(t\) leave \(k\) connected components, an MST must use exactly \(k-1\) edges of weight at least \(t\) to connect them. This observation gives the following characterization of MST edge weights.

\begin{lemma}
\label{lem:thresholds}
Let $G-R$ be connected, and let $T$ be any MST of $G-R$. For every
$t\ge0$,
\begin{align}
|\{e\in T:w_e\ge t\}|
&=\kappa(G_{<t}-R)-1,
\label{eq:threshold-strict}\\
|\{e\in T:w_e>t\}|
&=\kappa(G_{\le t}-R)-1.
\label{eq:threshold-closed}
\end{align}
All MSTs of $G-R$ have the same list of edge weights when arranged
in nondecreasing order. Moreover, a tree produced by Kruskal's algorithm
for the original weights remains an MST after each weight $w_e$ is
capped at $t$. The resulting MST value satisfies
\begin{equation}
\label{eq:threshold-integral}
F_t(R)=\int_0^t\bigl(\kappa(G_{<s}-R)-1\bigr)\,ds.
\end{equation}
\end{lemma}

\begin{proof}
Fix $t\ge0$, and let $k=\kappa(G_{<t}-R)$. Let $T_{<t}$ be the
forest consisting of all vertices and the edges of $T$ with weight
less than $t$. We claim that $T_{<t}$ connects all vertices within
each component of $G_{<t}-R$.
Suppose otherwise.  Since each component of $G_{<t}-R$ is connected
by edges of weight less than $t$, one of them contains an edge $e$
joining two different components of $T_{<t}$. Necessarily,
$e\notin T$; otherwise, it would belong to $T_{<t}$ and connect
these components. The unique path in $T$
between the endpoints of $e$ must contain an edge $f$ of weight
at least $t$. Adding $e$ to $T$ creates a cycle containing this
path, so replacing $f$ with $e$ gives a strictly lighter spanning
tree, contradicting the optimality of $T$.

Thus, the tree edges of weight less than $t$ form a spanning tree
within each of the $k$ components and number $n-k$ in total.
Since $T$ has $n-1$ edges, exactly $k-1$ of its edges have weight
at least $t$, proving~\eqref{eq:threshold-strict}.
The same argument using edges of weight at most $t$
proves~\eqref{eq:threshold-closed}.

Subtracting the two identities determines the number of tree edges
of weight exactly $t$ solely from $G-R$. Hence all MSTs have the same
number of edges of each weight, and therefore the same sorted list
of edge weights.

Capping preserves the nondecreasing edge order used by Kruskal's
algorithm~\cite{Kru56}. Running the algorithm in exactly this order
after capping therefore selects the same tree, which is optimal for
both weight functions.

Finally, take $T$ to be such a Kruskal MST. For each edge $e\in T$,
the values $s\in[0,t]$ satisfying $s\le w_e$ form an interval of
length $\min\{w_e,t\}$. Consequently,
\[
\begin{aligned}
F_t(R)
=\sum_{e\in T}\min\{w_e,t\}
=\int_0^t |\{e\in T:w_e\ge s\}|\,ds
=\int_0^t\bigl(\kappa(G_{<s}-R)-1\bigr)\,ds,
\end{aligned}
\]
where the last equality uses~\eqref{eq:threshold-strict}.
\end{proof}

We will use two consequences of Lemma~\ref{lem:thresholds}. First,
if a connected graph has all edge weights at most $t$, adding edges
of weight exactly $t$ does not change its MST value. Second, after
capping the weights at $t$, we may add artificial edges of weight
$t$ forming a spanning tree and forbid their deletion. These
protected edges keep the augmented graph connected under every
deletion of original edges, regardless of their total deletion cost. Whenever $G-R$ is connected, the
first observation ensures that adding these edges leaves its
capped MST value $F_t(R)$ unchanged.

For an undirected graph $H$ on $V$ with nonnegative edge costs, let $\delta_H(\mathcal P)$ denote the edges joining different parts of a partition $\mathcal P$ of $V$, and let $\delta_H(A)$ denote
the edges with exactly one endpoint in $A\subseteq V$. A $p$-cut is $\delta_H(\mathcal P)$ for a partition into $p$ nonempty
parts. The minimum-$p$-cut problem seeks the cheapest such set, equivalently, the cheapest edge set whose removal leaves $H$ with at least $p$ connected components. We use deterministic minimum-$p$-cut algorithms running in $N^{O(p)}$
time~\cite{GH94,CQX20}.

\subsection{Enumerating affordable partitions}
\label{sec:enumeration}

Both algorithms need to list affordable ways to divide a graph
into a bounded number of parts. The theorem below shows that this
is possible whenever the budget cannot create more than $k$
components. It builds on the tree-packing and cut-enumeration
framework of~\cite{CQX20}; Appendix~\ref{app:packing} gives the proof.
An enumerated partition need not have connected parts.

\begin{theorem}
\label{thm:partitions}
Let $H$ be an undirected multigraph on $n$ vertices, with nonnegative
rational edge costs $c_e$, and let $C\ge0$ be a rational budget. Fix an
integer $1\le k<n$. Suppose every partition of $V(H)$ into $k+1$ nonempty
parts has crossing cost greater than $C$. Then all partitions $\mathcal P$
satisfying
\[
  c(\delta_H(\mathcal P))\le C
\]
can be listed deterministically in
$n^{O(k)}\operatorname{poly}(N)$ time, where $N$ is the binary encoding
length of the input. Every such partition has at most $k$ parts.
The graph may be disconnected; loops can be ignored.
\end{theorem}

The proof packs spanning forests with exactly $k$ components. When
$C>0$, the hypothesis guarantees a packing of total weight at least
$C/(2k)$. Consequently, every affordable partition crosses at most
$2k$ edges of some forest in a polynomial-size support. Removing those
edges leaves at most $3k$ forest components, whose groupings recover the
partition. The case $C=0$ is handled directly by grouping the components
formed by positive-cost edges. This formulation also yields the
near-minimum-cut enumeration used by the two-threshold algorithm.

\begin{corollary}
\label{cor:nearcut}
Let $H$ be an undirected multigraph on $n$ vertices with nonnegative
rational edge costs. Let $2\le p\le n$, let $\lambda>0$ be its minimum
$p$-cut value, and let $\alpha\ge1$ be rational. All partitions into
exactly $p$ nonempty parts with crossing cost at most $\alpha\lambda$
can be listed deterministically in
$n^{O(\alpha p)}\operatorname{poly}(N)$ time.
\end{corollary}

\begin{proof}
Put $C=\alpha\lambda$ and $K=\lfloor2\alpha(p-1)\rfloor$.
If $n\le K$, enumerate all $p$-partitions directly. There are at most
$p^n\le n^n\le n^K$ of them, within the stated bound.

Suppose $n\ge K+1$. If a partition into $r=K+1$ parts had crossing
cost at most $C$, contract its parts and consider the resulting
multigraph, retaining only edges between distinct contracted vertices.
The sum of its weighted degrees is at most $2C$. Choose the $p-1$
vertices of smallest weighted degree, make each a separate part, and
group the remaining vertices into one part. The last part is nonempty
because $r>2\alpha(p-1)\ge p-1$. This gives a $p$-cut of cost at most
\[
  \frac{2(p-1)C}{r}
  =\frac{2\alpha(p-1)}{K+1}\lambda
  <\lambda,
\]
a contradiction. Thus every $(K+1)$-cut costs more than $C$.
Apply Theorem~\ref{thm:partitions} with parameter $K$ and budget $C$,
and retain exactly the partitions with $p$ parts. Its running time is
$n^{O(K)}\operatorname{poly}(N)
=n^{O(\alpha p)}\operatorname{poly}(N)$.
If $\lambda$ is not supplied, a deterministic minimum-$p$-cut algorithm
with running time $n^{O(p)}\operatorname{poly}(N)$ computes it within
the same bound~\cite{CQX20}.
\end{proof}

The positivity assumption on $\lambda$ is necessary for this enumeration
statement: even for fixed $p$, a graph may have exponentially many
zero-cost $p$-partitions. In the two-threshold algorithm, the corollary
is invoked only for a cut whose minimum cost is strictly positive.

\begin{revision}
\subsection{Residual rounding}
\label{subsec:residual-bound}

We use the component hierarchy and tree-knapsack rounding of
Linhares and Swamy~\cite{LS17} in the following form. The graph $H$
represents the residual instance after an initial deletion; protected
edges keep it connected under every subset of the deletable edges.
The bound retains the MST value already present before further
deletions. Section~\ref{subsec:rounding-proof} gives the proof.

\begin{lemma}
\label{lem:baseline}
Let $H$ have edge weights in $[0,t]$, and let $J$ be its set of
deletable edges, with positive deletion costs. Suppose $H-J$ is
connected, and write $h(S)=\MST(H-S)$ for $S\subseteq J$.
For a budget $D\ge0$, define
\begin{equation}
\label{eq:round-lagrangian}
L(D)=\min_{\lambda\ge0}
\left\{\lambda D+
\max_{S\subseteq J}\bigl(h(S)-\lambda c(S)\bigr)\right\}.
\end{equation}
One can compute $L(D)$ and a deletion $S\subseteq J$ in polynomial
time such that
\begin{equation}
\label{eq:round-baseline}
c(S)\le D,
\qquad
h(S)\ge\frac{L(D)+h(\varnothing)}2-t.
\end{equation}
\end{lemma}

For a fixed $\lambda$, the inner maximization assigns a penalty
$\lambda c(S)$ to a deletion. It ranges over all subsets of $J$,
including those that exceed $D$. For a feasible $Q$, the expression
being minimized is at least
$h(Q)+\lambda(D-c(Q))\ge h(Q)$, so $L(D)$ is an upper bound on the
budget-feasible optimum. The condition $H-J$ connected keeps every
value in this relaxation finite. Protected edges are excluded from $J$,
but an original deletable edge is not excluded merely because its
cost exceeds $D$.

The second algorithm needs only the following weaker consequence, obtained by using $L(D)\ge\OPT_H$ and $h(\varnothing)\ge0$ in
Lemma~\ref{lem:baseline}.

\begin{corollary}
\label{cor:additive}
Under the hypotheses of Lemma~\ref{lem:baseline}, let
$\OPT_H=\max\{h(S):S\subseteq J,\ c(S)\le D\}$.
A feasible deletion $S$ satisfying
\(
h(S)\ge\frac{\OPT_H}{2}-t
\)
can be found in polynomial time.
\end{corollary}
\end{revision}

\begin{revision}
\section{The Two-Threshold Algorithm}
\label{sec:approximation}

The algorithm combines cut enumeration with the residual rounding
bound of Lemma~\ref{lem:baseline}. Cut enumeration supplies the initial
deletions, and rounding uses the remaining budget to extend them.
We first give the complete procedure, then describe the enumeration
and prove the approximation guarantee.
\subsection{Building and comparing candidates}
\label{subsec:main-algorithm}

Algorithm~\ref{alg:main} first tests cuts at single thresholds.
These cuts and the empty deletion suffice when an optimal remaining
MST has at most one positive-weight edge. The main loop then tries
pairs of input weights. For a pair $a\ge t>0$, the procedure in
Section~\ref{subsec:initial-deletions} generates affordable initial
deletions $C$ that satisfy
\begin{equation}
\label{eq:two-requirements}
  \kappa(G_{<a}-C)\ge2,
  \qquad
  \kappa(G_{<t}-C)\ge3.
\end{equation}

For each such $C$, let $D=B-c(C)$ be the remaining budget.
If $D=0$, keep $C$ itself. Otherwise cap the edge weights of $G-C$
at $t$ and add a protected spanning tree of weight-$t$ edges to form
$H$. The rounding step may delete only the surviving original edges
of weight below $t$. Its deletable set and objective are
\begin{equation}
\label{eq:main-residual}
  J=\{e\in E\setminus C:w_e<t\},
  \qquad h(S)=\MST(H-S)\quad(S\subseteq J).
\end{equation}

\begin{algorithm}[H]
\caption{The two-threshold algorithm}
\label{alg:main}
\begin{minipage}{\linewidth}\editcolor
\textbf{Input:} A preprocessed instance with $n\ge2$, positive deletion
costs, budget $B>0$, finite optimum, and rational $0<\epsilon\le1$.\par
\textbf{Output:} A deletion $R$ with $c(R)\le B$.
\begin{enumerate}
\item Set $\eta=\epsilon/(2+\epsilon)$ and initialize the candidate
list    $\mathcal R=\{\varnothing\}$.
\item For each positive edge weight $a$, compute a minimum $2$-cut
$Q$ of $G_{<a}$. If $c(Q)\le B$, add $Q$ to $\mathcal R$.
After testing all such cuts, if $n=2$, return a candidate maximizing $F$.
\item For each pair of input edge weights $a\ge t>0$, run the procedure
of Lemma~\ref{lem:completion}, using $C=P\cup Z$ for each output pair.
For each initial deletion $C$ it produces, do the following:
\begin{enumerate}
\item Set $D=B-c(C)$. If $D=0$, add $C$ to
$\mathcal R$ and continue with the next initial deletion.
\item Form $(H,J,h)$ as in~\eqref{eq:main-residual}.
Apply Lemma~\ref{lem:baseline} with budget $D$ to obtain
$S\subseteq J$ with $c(S)\le D$.
\item Add $C\cup S$ to $\mathcal R$.
\end{enumerate}
\item Return a candidate $R\in\mathcal R$ maximizing
$F(R)=\MST_w(G-R)$ under the original weights.
\end{enumerate}
\end{minipage}
\end{algorithm}

The protected tree keeps $H-J$ connected. Its edges all have weight
$t$, so deleting other edges of capped weight $t$ cannot increase
the capped MST value. Every surviving original
edge of weight below $t$ belongs to $J$, even if its deletion cost
exceeds $D$.

For every feasible residual deletion $S$, the original deletion
$C\cup S$ is feasible and leaves $G$ connected. Consequently,
\begin{equation}
\label{eq:main-feasible-cap}
  h(S)=F_t(C\cup S)
  \qquad\text{whenever }S\subseteq J\text{ and }c(S)\le D.
\end{equation}
Protected edges are needed beyond these feasible deletions: they keep
$h$ finite on \emph{all} subsets of $J$, as required by the residual
relaxation in Lemma~\ref{lem:baseline}.

\subsection{Generating the initial deletions}
\label{subsec:initial-deletions}
\label{subsec:completion-proof}
For a fixed pair $(a,t)$ of edge weights, the procedure builds a set $P$ by
successive cuts and completes it with a set $Z$. It returns each
union $C=P\cup Z$ to Algorithm~\ref{alg:main}. The following lemma
also explains why the list contains a useful candidate whenever an
optimal deletion satisfies both threshold conditions.

\begin{lemma}
\label{lem:completion}
Let $n\ge3$, $a\ge t>0$, and let $0<\eta<1$ be rational. In
$n^{O(1/\eta)}\operatorname{poly}(N)$ time, one can enumerate pairs
$(P,Z)$ satisfying
\[
  P\cap Z=\varnothing,\qquad
  c(P)\le B,\qquad
  c(Z)\le\eta\bigl(B-c(P)\bigr),
\]
such that $C=P\cup Z$ satisfies~\eqref{eq:two-requirements}.
For every feasible deletion $R$ satisfying \eqref{eq:two-requirements},
some enumerated pair has $P\subseteq R$.
\end{lemma}

\begin{proof}
Start with $P=\varnothing$ and recurse. At a node, write
$D_0=B-c(P)$. For each condition not yet satisfied by $P$, compute
a cheapest cut that satisfies it: a minimum $2$-cut in $G_{<a}-P$
or a minimum $3$-cut in $G_{<t}-P$.

If any required cut costs more than $D_0$, discard the node.
If the sum of the required cut costs is at most $\eta D_0$, let $Z$
be their union and output $(P,Z)$. The union satisfies both conditions,
and its cost is at most the sum. All these cuts avoid $P$.
When both conditions already hold, this rule outputs
$(P,\varnothing)$.

Otherwise, one required minimum cut has cost
$\lambda>\eta D_0/2$. Let $p\in\{2,3\}$ be the required number
of parts, and set $K=G_{<a}-P$ if $p=2$ and $K=G_{<t}-P$ if $p=3$.  Since the node was not
discarded,
\[
  0<\lambda\le D_0<(2/\eta)\lambda.
\]
Corollary~\ref{cor:nearcut} lists all $p$-partitions of $K$ with
crossing cost at most $(2/\eta)\lambda$. For each corresponding cut
$Q$ with $c(Q)\le D_0$, recurse with $P\cup Q$.
The cut $Q$ avoids $P$ and permanently satisfies an outstanding
condition. Deletions preserve conditions already satisfied, so there
are at most two branching steps.

To prove completeness, fix a feasible $R$ satisfying both conditions.
Follow nodes with $P\subseteq R$, starting at the root.
The remaining deletion $R\setminus P$ costs at most $D_0$ and
satisfies every outstanding condition, so such a node is not discarded.
At a branching step, group the components of $K-(R\setminus P)$
into exactly $p$ nonempty parts. Their crossing set $Q$ in $K$ is
contained in $R\setminus P$ and costs at most $D_0$.
It is therefore enumerated, and the associated child still satisfies
$P\subseteq R$. Following these children reaches an output pair.
The final $Z$ may use edges outside $R$, but has the required cost.

Every output is feasible because
$c(P\cup Z)\le c(P)+\eta(B-c(P))\le B$.
The cut enumeration takes $n^{O(1/\eta)}\operatorname{poly}(N)$ time
per branching node and has at most that many outputs.
Depth at most two gives the stated total running time.
\end{proof}

\subsection{Approximation guarantee}
\label{subsec:main-guarantee}
\label{subsec:rounding}

\begin{proof}[Proof of Theorem~\ref{thm:approximation}]
First assume that $0<\epsilon\le1$ is rational and put
$\eta=\epsilon/(2+\epsilon)$.
Every candidate in Algorithm~\ref{alg:main} respects the budget.
Fix an optimal deletion $R^\star$ and an MST of $G-R^\star$.
If its edges all have weight zero, then $\OPT=0$ and the guarantee is immediate.
If its only positive edge has weight $a$, then $\OPT=a$ and $G_{<a}-R^\star$ is disconnected by Lemma~\ref{lem:thresholds}. A minimum $2$-cut $Q$ of $G_{<a}$ therefore has cost at most $c(R^\star)\le B$. Since $G_{<a}-Q$ is disconnected, every spanning tree of $G-Q$ uses an edge of weight at least $a$.
Thus $F(Q)\ge a=\OPT$, and a preliminary candidate is optimal. This also covers $n=2$.

Otherwise, let $a\ge t>0$ be the two largest edge weights in this
MST. Their pair is enumerated, and $R^\star$ satisfies both
conditions in~\eqref{eq:two-requirements}. By
Lemma~\ref{lem:completion}, some output $(P,Z)$ has
$P\subseteq R^\star$. Fix this output and write
\[
  C=P\cup Z,\qquad D_0=B-c(P),\qquad D=B-c(C).
\]

Table~\ref{tab:threshold-roles} summarizes how the two conditions on $C$ enter the proof.

\begin{table}[ht]
\editcolor
\caption{How the two threshold conditions control the losses.}
\label{tab:threshold-roles}
\begin{tabularx}{\linewidth}{
  @{}l@{\hspace{1.5em}}
  >{\hsize=.8\hsize}X@{\hspace{1em}}
  >{\hsize=1.2\hsize}X@{}
}
\toprule
Threshold & Required MST edges & Effect on the guarantee\\
\midrule
$t$ & Two edges of weight at least $t$
& $F_t(C)\ge2t$ cancels the additive rounding loss $t$\\
$a$ & One edge of weight at least $a$
& Restoring the weights adds at least $a-t$\\
\bottomrule
\end{tabularx}
\end{table}

\paragraph{Remaining budget and initial value.} Since $P\cap Z=\varnothing$, the completion guarantee gives
\begin{equation}
\label{eq:remaining-budget}
  D=D_0-c(Z)\ge(1-\eta)D_0.
\end{equation}
If $D=0$, then $D_0=0$. Positive deletion costs imply
$Z=\varnothing$ and, because $P\subseteq R^\star$, $P=R^\star$.
Thus the candidate $C$ is optimal. Assume henceforth that $D>0$.
Construct the residual instance $(H,J,h)$ as above.
The condition at $t$ and~\eqref{eq:main-feasible-cap} give
\begin{equation}
\label{eq:initial-two-t}
  h(\varnothing)=F_t(C)\ge2t.
\end{equation}

\paragraph{A comparison at the reduced budget.}
The remaining part of the optimal deletion gives the witness
\[
  S^\star=\{e\in R^\star\setminus C:w_e<t\}\subseteq J.
\]

Its cost satisfies $c(S^\star)\le D_0$. For every threshold $s<t$, the deletion $C\cup S^\star$ removes every edge of $R^\star$ with weight below $s$, and possibly additional edges. Protected edges are absent from both threshold graphs.  The threshold integral therefore
implies
\begin{equation}
\label{eq:comparison-cap}
  h(S^\star)\ge F_t(R^\star)=\OPT-a+t.
\end{equation}
The equality holds because only the largest edge of the comparison
MST can exceed $t$. This argument uses the connected auxiliary graph;
it does not require the possibly over-budget graph
$G-(C\cup S^\star)$ to be connected.

Let $L(D)$ be the residual relaxation of Lemma~\ref{lem:baseline}. The completion guarantee and $c(S^\star)\le D_0$ give
$(1-\eta)c(S^\star)\le(1-\eta)D_0\le D$. Thus the convex combination with weights $1-\eta$ and $\eta$ on $S^\star$ and $\varnothing$ has average cost at most $D$, even if $S^\star$ itself exceeds this budget. For each $\lambda\ge0$, its inner maximum is at least the convex combination of the values at $S^\star$ and $\varnothing$. Hence
\begin{align*}
  &\lambda D+\max_{Q\subseteq J}\{h(Q)-\lambda c(Q)\}\\
  &\quad\ge (1-\eta)h(S^\star)+\eta h(\varnothing)
       +\lambda\bigl(D-(1-\eta)c(S^\star)\bigr)\\
  &\quad\ge (1-\eta)h(S^\star).
\end{align*}
The last inequality uses $h(\varnothing)\ge0$ and $D-(1-\eta)c(S^\star)\ge0$.

Minimizing over $\lambda$ and using~\eqref{eq:comparison-cap} gives
\begin{equation}
\label{eq:lagrangian-mixture}
  L(D)\ge(1-\eta)(\OPT-a+t).
\end{equation}
This is why the relaxation ranges over all subsets of $J$: the
comparison witness remains available even when its cost exceeds $D$.
The convex combination is only an analysis device; the algorithm is
deterministic.

\paragraph{Rounding and restoring the weights.}
Lemma~\ref{lem:baseline} returns a deletion $S\subseteq J$ of cost
at most $D$. The initial value in~\eqref{eq:initial-two-t} cancels
the additive rounding loss:
\begin{equation}
\label{eq:rounded-residual}
  h(S)\ge\frac{L(D)+h(\varnothing)}2-t\ge\frac{L(D)}2.
\end{equation}
The condition at $a$ persists after deleting $S$. A Kruskal MST of
$G-(C\cup S)$ therefore contains an edge of weight at least $a$,
and remains optimal after capping at $t$
(Lemma~\ref{lem:thresholds}). Restoring the original weights adds
at least $a-t$. Thus
\begin{align}
  F(C\cup S)
  &\ge h(S)+a-t \nonumber\\
  &\ge\frac{1-\eta}{2}(\OPT-a+t)+a-t \nonumber\\
  &=\frac{1-\eta}{2}\OPT+\frac{1+\eta}{2}(a-t)
   \ge\frac{\OPT}{2+\epsilon}.
\label{eq:restored-guarantee}
\end{align}
The algorithm returns a candidate at least as good as this one.

There are at most $m^2$ threshold pairs. For each pair, completion
takes $n^{O(1/\eta)}\operatorname{poly}(N)$ time, including all its
outputs, and each residual rounding takes polynomial time.
Since $1/\eta=1+2/\epsilon$, the total running time is
$N^{O(1/\epsilon)}$. For a nonrational $0<\epsilon\le1$, use a
rational parameter between $\epsilon/2$ and $\epsilon$; both the
approximation guarantee and the asymptotic running time are preserved.
For $\epsilon>1$, use parameter $1$.
\end{proof}
\end{revision}

\subsection{\edit{Residual rounding bound: Proof of Lemma~\ref{lem:baseline}}}
\label{subsec:rounding-proof}

\edit{It remains to prove Lemma~\ref{lem:baseline}, which supplies the
rounding step used by both algorithms. Section~\ref{app:relaxation}
gives the exact optimization and encoding details for its Lagrangian
relaxation.}

The rounding argument starts with two nested deletions, a feasible
one and an over-budget one. A tree records the additional components
created by the larger deletion at successive weight levels. We select
some of these components without exceeding the budget. The factor
$1/2$ comes from counting the cost of an edge at most twice, and the
additive loss comes from one path in this tree.

\begin{proof}[Proof of Lemma~\ref{lem:baseline}]
The cases $t=0$, $|V(H)|\le1$, or $J=\varnothing$ are immediate.
Section~\ref{app:relaxation} gives an exact polynomial-time procedure
that computes $L(D)$ and either returns a feasible set of value
$L(D)$ or finds two maximizers $S_-\subseteq S_+$ of
$h(S)-\lambda^*c(S)$ at an optimal multiplier $\lambda^*$, with
\[
c(S_-)<D<c(S_+).
\]
A feasible set of value $L(D)$ already satisfies the lemma, since
$L(D)\ge h(\varnothing)$. \edit{This outcome also covers $D=0$.} In the second case, put
\[
\theta=\frac{D-c(S_-)}{c(S_+)-c(S_-)}\in(0,1).
\]
The two sets have the same penalized value, so
\begin{equation}
\label{eq:round-interpolation}
L(D)=(1-\theta)h(S_-)+\theta h(S_+).
\end{equation}
The task is to retain $h(S_-)$ and obtain part of the increase
$h(S_+)-h(S_-)$ using the budget $D-c(S_-)$.

\paragraph{\edit{Components, costs, and profits.}}
List the distinct positive edge weights below $t$, together with
the endpoints, as $0=t_0<t_1<\cdots<t_\ell=t$.
For every $i<\ell$, create a node $(A,i)$ for each component $A$ of
$H_{\le t_i}-S_+$. Its parent is the component containing it at the
next level; at level $\ell$, use a single root for $V(H)$.
Components can merge as the threshold increases, but they cannot
split. They therefore form a rooted tree. Equal vertex sets at
different levels give different nodes.

For a nonroot node $v=(A,i)$, let
\[
D_v=\{e\in\delta_H(A)\setminus S_-:w_e=t_i\},
\qquad d_v=c(D_v).
\]
These edges all belong to $S_+\setminus S_-$: otherwise an edge
of weight $t_i$ would leave a component of $H_{\le t_i}-S_+$.
An edge can occur only at its own weight level and in the two nodes
containing its endpoints. Hence
\begin{equation}
\label{eq:round-costs}
\sum_v d_v\le2\bigl(c(S_+)-c(S_-)\bigr).
\end{equation}
\edit{Counting an edge at both endpoints only overestimates the cost
of the union of the selected deletion sets. This double counting
explains the factor $1/2$ in the fractional selection below.}

We assign profit only to the components gained beyond $S_-$.
Within each component of $H_{\le t_i}-S_-$, choose one of its
$S_+$ components as a representative and give that node profit zero.
Give every other node at level $i$ profit $p_v=t_{i+1}-t_i$.
The representative contributes no new profit because the component
containing it is already counted in $h(S_-)$. For example, if one
component splits into three, the component count increases by two,
so only two nodes receive profit.

\edit{Representatives at different levels need not be consistent,
because the gain is measured separately at each level. For
$s\in(t_i,t_{i+1})$, the graphs $H_{<s}$ and $H_{\le t_i}$ coincide.}
The threshold integral gives
\begin{equation}
\label{eq:round-integral}
h(Q)=\int_0^t\bigl(\kappa(H_{<s}-Q)-1\bigr)\,ds
\qquad(Q\subseteq J).
\end{equation}
Consequently,
\begin{equation}
\label{eq:round-profits}
\sum_v p_v=h(S_+)-h(S_-).
\end{equation}
Every root-to-leaf path contains at most one node at each level, so
its total profit is at most $t$.

\paragraph{\edit{Selecting components within the budget.}}
Selecting $(A,i)$ means isolating $A$ at threshold $t_i$.
Deleting only $D_v$ removes boundary edges of weight exactly $t_i$.
We must also remove lighter boundary edges; their endpoint components
appear among the descendants of $(A,i)$. We therefore require that
every selected node have all its descendants selected.

Lemma~\ref{lem:tree} rounds a fractional selection of this kind with
a loss of at most the largest root-to-leaf profit. The constant
selection $x_v=\theta/2$ satisfies its nesting constraints, and
\eqref{eq:round-costs} gives
\[
\sum_v d_vx_v
\le\theta\bigl(c(S_+)-c(S_-)\bigr)=D-c(S_-).
\]
Applying the lemma with this remaining budget yields a selection
$X$ containing every descendant of each selected node, with
\[
\sum_{v\in X}d_v\le D-c(S_-),\qquad
\sum_{v\in X}p_v
\ge\frac{\theta}{2}\bigl(h(S_+)-h(S_-)\bigr)-t.
\]
Return $S=S_-\cup\bigcup_{v\in X}D_v$. Then $S\subseteq S_+$
and $c(S)\le D$.

\paragraph{\edit{Recovering the gain over $S_-$.}}
If $(A,i)$ is selected, every edge leaving $A$ of weight at most
$t_i$ is in $S_-$ or in $D_v$ for that node or one of its
descendants. All such edges are therefore deleted. Also, $A$ remains
connected at that level: it was connected after deleting the larger
set $S_+$. Thus $A$ is a component of $H_{\le t_i}-S$.

\edit{Fix a level $i$. Within each component of $H_{\le t_i}-S_-$,
the selected nonrepresentative nodes give distinct components of
$H_{\le t_i}-S$. The representative component lies in a further
component, distinct from all these selected components. Thus the
component count increases by at least the number of selected
nonrepresentative nodes.} Integrating over the weight levels gives
\[
\begin{aligned}
h(S)
&\ge h(S_-)+\sum_{v\in X}p_v\\
&\ge h(S_-)+\frac{\theta}{2}\bigl(h(S_+)-h(S_-)\bigr)-t\\
&=\frac{L(D)+h(S_-)}2-t
\ge\frac{L(D)+h(\varnothing)}2-t.
\end{aligned}
\]
\edit{The analysis retains the entire value of $S_-$ and approximates
only the additional component value. Monotonicity then allows
$h(S_-)$ to be replaced by $h(\varnothing)$ in the stated bound.}
The component tree has polynomial size. Its construction and the
rounding procedure are polynomial-time, as are the exact optimization
steps in Section~\ref{app:relaxation}.
\end{proof}

\section{\edit{The Heavy Quotient Tree Algorithm}}
\label{sec:quotient}

The second algorithm separates the heavy edges connecting
different parts of the graph from the lighter edges within
those parts. It chooses a threshold $q$ that bounds the
number of light-edge components after every feasible
deletion, and then enumerates their possible partitions
and the heavy trees connecting them. For each such choice,
the remaining optimization reduces to a bounded-weight
MST-interdiction instance, to which
Corollary~\ref{cor:additive} applies.
The algorithm returns the better of a threshold-cut solution
and a solution obtained from the auxiliary instances.
The former offsets the additive loss in the latter,
yielding the following approximation guarantee.

\begin{theorem}
\label{thm:quotient}
For every fixed integer $k\ge1$ and every instance with $n\ge k+1$,
there is a deterministic $(2+2/k)$-approximation for MST interdiction
running in $N^{O(k)}$ time and respecting the deletion budget.
\end{theorem}

\subsection{Algorithm}
\label{subsec:quotient-algorithm}

For a feasible deletion $R$, let
$x_1(R)\ge\cdots\ge x_{n-1}(R)$ be the edge weights
of a remaining MST, in nonincreasing order.
These weights are independent of the chosen MST by
Lemma~\ref{lem:thresholds}.
Fix an integer $1\le k\le n-1$.
Define $q=\max_{c(R)\le B}x_k(R)$ and $H=G_{\le q}$.
The following lemma establishes two properties of this threshold.

\begin{lemma}
\label{lem:quotient-threshold}
The threshold $q$ and a feasible interdiction set $R_q$ can be computed in $N^{O(k)}$ time such that
\begin{enumerate}
    \item $F(R_q)\ge kq$, and
    \item $\kappa(H-R)\le k$ for every feasible $R$. In particular, every $(k+1)$-cut of $H$ costs more than $B$.
\end{enumerate}
\end{lemma}

\begin{proof}
For every feasible $R$ and every $t\ge0$, Lemma~\ref{lem:thresholds} gives
\[
\begin{aligned}
x_k(R)\ge t
&\iff \kappa(G_{<t}-R)\ge k+1,\\
x_k(R)>t
&\iff \kappa(G_{\le t}-R)\ge k+1.
\end{aligned}
\]
By the first equivalence, there exists a feasible deletion
$R$ with $x_k(R)\ge t$ if and only if the minimum
$(k+1)$-cut of $G_{<t}$ costs at most $B$.
We test all candidate thresholds $t$ among zero and the
distinct input edge weights, and let $q$ be the largest
threshold passing this test.
The edges of a corresponding minimum cut in $G_{<q}$
form a feasible interdiction set $R_q$ with
$x_k(R_q)\ge q$, and hence $F(R_q)\ge kq$.

If some feasible $R$ satisfies
$\kappa(H-R)\ge k+1$, the second equivalence gives
$x_k(R)>q$, contradicting the definition of $q$.
Therefore $\kappa(H-R)\le k$ for every feasible $R$.
Equivalently, every $(k+1)$-cut of $H$ costs more
than $B$.
\end{proof}

The first inequality of Lemma~\ref{lem:quotient-threshold} provides a candidate solution of value at least $kq$;
the second makes it possible to enumerate the light-edge partitions.

\paragraph{Enumeration of light-edge partitions and heavy quotient trees.}
By Lemma~\ref{lem:quotient-threshold} and Theorem~\ref{thm:partitions}, 
we can enumerate, in $n^{O(k)}\operatorname{poly}(N)$ time, every vertex partition
$\mathcal P=\{P_1,\ldots,P_r\}$ satisfying $c(\delta_H(\mathcal P))\le B$ and $r\le k$.
\edit{An enumerated part need not be connected. On the branch used
for comparison with an optimal deletion, the parts are the connected
components of the remaining light-edge subgraph.}
Let $G/\mathcal P$ denote the quotient multigraph obtained by
contracting each part into a single vertex, discarding
self-loops, and retaining parallel edges with their original
weights. For $R\subseteq E$, define $(G-R)/\mathcal P$
analogously.

For each enumerated partition $\mathcal P=\{P_1,\ldots,P_r\}$, we enumerate all spanning trees \(T\) of \(G/\mathcal P\) whose edges have weight greater than \(q\).
Each $T$ consists of $r-1$ edges, so there are at most $m^{r-1}\le m^{k-1}$ choices.
\edit{For a one-part partition, use the empty tree $T=\varnothing$.}
Write $w(T)=\sum_{e\in T}w_e$.

\edit{Recall that the cycle optimality condition for MSTs states that a
spanning tree is minimum if and only if every non-tree edge has weight
at least the maximum edge weight on its tree path.}
For each guessed tree $T$, we delete the inter-part edges that
would violate its MST cycle optimality condition. For distinct
parts $P_i,P_j$, let $\lambda_T(i,j)$ be the maximum edge weight
on their unique path in $T$, and define the \emph{canonical deletion}
\begin{equation}
\label{eq:quotient-canonical}
R_0(\mathcal P,T)=
\{uv\in E:u\in P_i,\ v\in P_j,\ i\ne j,\
w_{uv}<\lambda_T(i,j)\}.
\end{equation}
\edit{The strict inequality retains edges tied with the largest weight
on their tree path. For a one-part partition, $R_0=\varnothing$.}

\begin{proposition}
\label{prop:canonical-quotient}
For every enumerated pair $(\mathcal P,T)$, all surviving
inter-part edges of $G-R_0$ have weight greater than $q$,
and $T$ is an MST of the quotient graph $(G-R_0)/\mathcal P$.
\end{proposition}

\begin{proof}
Every edge $e\in T$ survives, since its weight equals
the maximum weight on its own $T$-path. For any
surviving inter-part edge $e$ joining $P_i$ and $P_j$,
we have $w_e\ge\lambda_T(i,j)>q$.
Thus $T$ is an MST of the quotient graph by the
cycle optimality condition.
\end{proof}


Discard a branch if $c(R_0)>B$; otherwise set
$B'=B-c(R_0)$.

\paragraph{The bounded-weight auxiliary instance.}
We now optimize deletions within the parts while keeping
the heavy connections \edit{fixed. For} each retained pair $(\mathcal P,T)$, construct an auxiliary graph $A$ on $V$ as follows:
\begin{enumerate}[label=(\roman*)]
\item Within each $P_i$, retain every original edge of weight at most
$q$, with its original deletion cost.  These edges form the deletable
set $J$.
\item Within each $P_i$, add a protected spanning tree whose edges have
weight $q$.  This keeps that part connected after any deletion from $J$.
\item Choose one representative in each part and connect the
representatives by a protected tree of zero-weight edges.  These edges
combine the internal problems without contributing to the MST weight.
\end{enumerate}
The protected edges together form a spanning tree of $A$, and every
edge weight lies in $[0,q]$.
\edit{All original edges specified in (i) remain in $J$, even if
their deletion costs exceed $B'$, because the residual relaxation
ranges over all subsets of $J$.}
Apply Corollary~\ref{cor:additive} to the auxiliary instance
with budget $B'$ to obtain a deletion $S\subseteq J$ satisfying
$c(S)\le B'$. The resulting deletion $R_0\cup S$ is feasible,
since $c(R_0\cup S)\le c(R_0)+B'=B$.

Algorithm~\ref{alg:quotient} summarizes the candidate construction and
returns the best feasible deletion under the original edge weights.

\begin{algorithm}[H]
\caption{MST interdiction via a heavy quotient tree}
\label{alg:quotient}
\begin{minipage}{\linewidth}
\textbf{Input:} A preprocessed instance $(G,w,c,B)$ with finite optimum
and an integer $1\le k\le n-1$.\par
\textbf{Output:} A deletion $R$ with $c(R)\le B$.
\begin{enumerate}
\item Compute $q$ and a feasible deletion $R_q$ using
Lemma~\ref{lem:quotient-threshold}. Set $H=G_{\le q}$ and initialize
$\mathcal R=\{R_q\}$.
\item Use Theorem~\ref{thm:partitions} to enumerate all partitions
$\mathcal P=\{P_1,\ldots,P_r\}$ with
$c(\delta_H(\mathcal P))\le B$.
\item For each such $\mathcal P$, enumerate all spanning
trees $T$ of $G/\mathcal P$ using only edges of weight
greater than $q$.
\edit{For a one-part partition, use $T=\varnothing$.
For each such $T$, construct $R_0=R_0(\mathcal P,T)$
and perform Steps~4 and~5.}
\item If $c(R_0)>B$, discard this branch.  Otherwise set
$B'=B-c(R_0)$ and construct the auxiliary graph $A(\mathcal P,T)$
and deletable set $J$.
\item Apply Corollary~\ref{cor:additive} to $(A,J,B')$ and obtain
$S\subseteq J$ with $c(S)\le B'$.  Add $R_0\cup S$ to $\mathcal R$.
\item Return a set $R\in\mathcal R$ maximizing the original MST
weight $F(R)=\MST_w(G-R)$.
\end{enumerate}
\end{minipage}
\end{algorithm}

\subsection{Approximation guarantees}
\label{subsec:quotient-analysis}

For a retained branch $(\mathcal P,T)$ and its auxiliary graph $A$, define
\[
\OPT_A=\max\{\MST(A-S):S\subseteq J,\ c(S)\le B'\}.
\]
By Corollary~\ref{cor:additive}, the algorithm finds
$S\subseteq J$ satisfying $c(S)\le B'$ and
\begin{equation}
\label{eq:quotient-oracle}
\MST(A-S)\ge\tfrac12\OPT_A-q.
\end{equation}

The following lemma relates the MST value of a feasible
interdiction set to that of the auxiliary instance.

\begin{lemma}
\label{lem:quotient-branch}
For every retained branch $(\mathcal P,T)$ and every
feasible auxiliary deletion $S\subseteq J$,
\begin{equation}
\label{eq:quotient-lifting}
F(R_0\cup S)\ge w(T)+\MST(A-S).
\end{equation}
\end{lemma}

\begin{proof}
Consider the graph obtained from $G-(R_0\cup S)$
by adding the protected weight-$q$ spanning trees
within the parts. Its MST value is at most
$F(R_0\cup S)$.
By Proposition~\ref{prop:canonical-quotient}, every surviving
inter-part edge has weight greater than $q$, and $T$ is an
MST of the quotient graph. Since $S$ contains only internal
edges, the quotient graph is unchanged. Kruskal's algorithm
therefore first constructs an MST within each part, using
edges of weight at most $q$, and then connects the parts
using $T$.

The internal MST values equal those in $A[P_i]-S$,
since internal edges heavier than $q$ cannot improve an MST
in the presence of a weight-$q$ spanning tree.
The protected zero-weight edges between parts are bridges
in $A-S$, so
\[
\MST(A-S)=\sum_{i=1}^r\MST(A[P_i]-S).
\]
Thus the augmented graph has MST weight
$w(T)+\MST(A-S)$, proving~\eqref{eq:quotient-lifting}.
\end{proof}

We next show that the enumeration contains a branch
that captures the structure and value of an optimal solution.

\begin{lemma}
\label{lem:quotient-optimal}
For every optimal interdiction set $R^\star$, the
enumeration contains a branch
$(\mathcal P^\star,T^\star)$ such that
\[
R_0^\star:=
R_0(\mathcal P^\star,T^\star)\subseteq R^\star.
\]
Moreover, the optimal value $\OPT_{A^\star}$ of
this branch's auxiliary instance satisfies
\begin{equation}
\label{eq:quotient-capture}
\OPT_{A^\star}\ge\OPT-w(T^\star).
\end{equation}
\end{lemma}

\begin{proof}
Fix an optimal deletion $R^\star$, and let $\mathcal P^\star$
be the partition into connected components of $H-R^\star$.
By Lemma~\ref{lem:quotient-threshold}, it has at most $k$
parts. Every edge of $H$ joining distinct parts belongs to
$R^\star$, so
\[
c(\delta_H(\mathcal P^\star))\le c(R^\star)\le B.
\]
Thus $\mathcal P^\star$ is enumerated.
Let $T^\star$ be an MST of $(G-R^\star)/\mathcal P^\star$.
Every edge of this quotient graph has weight greater than
$q$, since otherwise its endpoints would lie in the same
component of $H-R^\star$. Hence $T^\star$ is also enumerated.
We claim that $R_0^\star\subseteq R^\star$.
Otherwise, some edge $e\in R_0^\star$ survives $R^\star$,
with weight strictly less than the maximum edge weight
on its $T^\star$-path. This contradicts the MST cycle
optimality condition for $T^\star$.
Thus $R_0^\star\subseteq R^\star$, and the branch is retained.

Let $A^\star$ and $J^\star$ be the auxiliary graph and
deletable edge set of this branch, and define
\[
S^\star=(R^\star\setminus R_0^\star)\cap J^\star.
\]
Since $R_0^\star\subseteq R^\star$,
\[
c(S^\star)\le B-c(R_0^\star)=B',
\]
so $S^\star$ is feasible for the auxiliary instance.

Each part $P_i^\star$ is connected in $H-R^\star$
using edges of weight at most $q$. Adding protected
weight-$q$ edges therefore does not change its
internal MST value. Consequently,
\[
\MST(A^\star-S^\star)
=\sum_{i=1}^r
\MST\bigl((H-R^\star)[P_i^\star]\bigr).
\]
Since every surviving inter-part edge has weight
greater than $q$, Kruskal's algorithm first
connects each part internally and then connects
the parts using a quotient MST of weight
$w(T^\star)$. Hence
\[
\OPT=F(R^\star)
=w(T^\star)+\MST(A^\star-S^\star).
\]
Since $S^\star$ is feasible for the auxiliary instance,
\[
\OPT_{A^\star}
\ge\MST(A^\star-S^\star)
=\OPT-w(T^\star),
\]
proving~\eqref{eq:quotient-capture}.
\end{proof}

Combining the two lemmas with the threshold-cut solution
yields the approximation guarantee.

\begin{proof}[Proof of Theorem~\ref{thm:quotient}]
Consider the branch $(\mathcal P^\star,T^\star)$
guaranteed by Lemma~\ref{lem:quotient-optimal}.
Let $S$ be the auxiliary deletion returned by
Corollary~\ref{cor:additive}, and set
$R_c=R_0^\star\cup S$.
By Lemmas~\ref{lem:quotient-branch} and
\ref{lem:quotient-optimal}, together with
\eqref{eq:quotient-oracle},
\[
\begin{aligned}
F(R_c)
&\ge w(T^\star)+\MST(A^\star-S)\\
&\ge w(T^\star)+\tfrac12\OPT_{A^\star}-q\\
&\ge w(T^\star)
   +\tfrac12\bigl(\OPT-w(T^\star)\bigr)-q\\
&\ge\tfrac12\OPT-q.
\end{aligned}
\]

Let $\mathrm{ALG}$ be the maximum MST value among
all candidates returned by the algorithm.
The candidate $R_c$ and the threshold-cut candidate
$R_q$ imply
\[
\mathrm{ALG}\ge\tfrac12\OPT-q,
\qquad
\mathrm{ALG}\ge F(R_q)\ge kq.
\]
Thus $q\le\mathrm{ALG}/k$, and
\[
\OPT
\le2\mathrm{ALG}+2q
\le\left(2+\frac2k\right)\mathrm{ALG}.
\]
This proves the approximation guarantee.

Computing $q$ requires minimum-$(k+1)$-cut
computations, taking $N^{O(k)}$ time.
Theorem~\ref{thm:partitions} enumerates
$n^{O(k)}\operatorname{poly}(N)$ partitions.
Each partition has at most $m^{k-1}$ heavy 
quotient trees, and constructing and solving each
auxiliary instance takes polynomial time.
The overall running time is therefore $N^{O(k)}$.
\end{proof}

For any $0<\epsilon\le1$, set $k=\lceil2/\epsilon\rceil$.
If $n\ge k+1$, Theorem~\ref{thm:quotient} yields a
$(2+\epsilon)$-approximation in $N^{O(1/\epsilon)}$ time.
Otherwise, $n\le k$, and the exact algorithm of
Lemma~\ref{lem:fixedn} solves the instance in
$N^{O(n)}=N^{O(1/\epsilon)}$ time.
For $\epsilon>1$, we use $\epsilon=1$, obtaining the
overall running-time bound $N^{O(1+1/\epsilon)}$.

\begin{revision}
\subsection{\edit{Exact Optimization on Small Instances}}
\label{app:small-instances}

The second algorithm uses a $k$th largest tree-edge weight, which is
defined only when $n\ge k+1$. This subsection handles $n\le k$ before
that threshold is computed. The enumeration fixes a possible remaining
MST and removes only edges that would prevent it from being optimal.

\begin{lemma}
\label{lem:fixedn}
MST interdiction on $n$ vertices can be solved exactly in $N^{O(n)}$
time. 
\end{lemma}

\begin{proof}
Return the empty deletion when $n=1$, and otherwise handle a feasible
disconnection first. Enumerate every set of $n-1$ original edges that forms a spanning
tree $T$. Let $P_T(u,v)$ be its unique $u$--$v$ path,
and put
\[
R_T=\{uv\in E:w_{uv}<\max_{e\in P_T(u,v)}w_e\}.
\]
No edge of $T$ lies in $R_T$. Every surviving non-tree edge has weight
at least the maximum on its tree path, so $T$ is an MST of $G-R_T$.
Thus each candidate with $c(R_T)\le B$ is feasible and has value
$\sum_{e\in T}w_e$. The strict inequality retains tied edges, which
is consistent with the non-strict cycle optimality condition.

Let $T^\star$ be an MST after an optimal deletion $R^\star$.
Every edge in $R_{T^\star}$ must belong to $R^\star$, since otherwise
it would violate cycle optimality in $G-R^\star$. Hence this candidate
is feasible and has value exactly $\OPT$. Returning the best feasible
candidate is therefore optimal. There are at most $m^{n-1}$ tree
candidates, with polynomial work per candidate, giving the claimed
running time.
\end{proof}

\end{revision}

\section{Hardness and the Metric-TSP Consequence}
\label{sec:consequences}

We conclude the main results with a conditional lower bound for MST
interdiction and a consequence for metric-TSP interdiction.
The lower bound explains why a factor below $2$ would require
overcoming SSEH. The consequence uses the relation between spanning
trees and covering closed walks.

\subsection{Conditional hardness}
\label{sec:hardness}

We derive the following lower bound from the Maximum Components
hardness of Aissi, Attias, Chen, and Ravi~\cite[Theorem~5]{AACR22}.

\paragraph{An objective-preserving transfer.}
Maximum Components asks for a deletion of cost at most $B$ maximizing
the number of remaining components of a graph $K=(U,F)$.
Give every edge of $K$ weight zero, keeping its deletion cost.
Add a new vertex $r$ and an edge $ru$ of weight one and deletion cost
$B+1$ for each $u\in U$; call the resulting graph $\widehat G$.
For every feasible deletion $R\subseteq F$,
\begin{equation}
\label{eq:components-transfer}
\MST(\widehat G-R)=\kappa(K-R).
\end{equation}
Indeed, each zero-weight component needs one star edge to reach $r$,
and one such edge per component suffices. Feasible deletions are
identical in the two instances, and the objective values agree
exactly, with no additive shift.

Thus a $(2-\epsilon)$-approximation for MST interdiction would
guarantee a fraction $1/(2-\epsilon)>1/2$ of the Maximum Components
optimum. When the source graph is simple, the construction is also simple,
has weights in $\{0,1\}$, and remains connected after every feasible
deletion.

\begin{theorem}
\label{thm:hardness}
Assume SSEH. For every fixed $0<\epsilon<1$, MST interdiction has no
deterministic polynomial-time $(2-\epsilon)$-approximation unless
$\mathsf{P}=\mathsf{NP}$. This holds even for simple graphs with edge
weights in $\{0,1\}$ and a spanning star whose edges each have deletion
cost greater than the budget.
\end{theorem}

Appendix~\ref{app:hardness} gives the proof.

\subsection{Metric-TSP interdiction}
\label{sec:tsp}

Following~\cite{Zen15,LS17}, consider a complete undirected input graph
with nonnegative metric weights. After a feasible deletion $R$, the
operator chooses a minimum-weight closed walk visiting every vertex of
$G-R$, allowing repeated edges and vertices. Denote its value by
$\TSP(G-R)$, with value $+\infty$ if $G-R$ is disconnected.
Equivalently, this is TSP in the shortest-path metric recomputed after
the deletion: metric edges can be expanded into surviving paths, and
a covering walk can be shortcut in that metric.

The support of a covering closed walk contains a spanning tree, while
doubling an MST gives a covering closed walk. Thus, for every connected
remaining graph,
\begin{equation}
\label{eq:tsp-bounds}
F(R)\le\TSP(G-R)\le2F(R).
\end{equation}

\begin{corollary}
\label{cor:tsp}
For every fixed $\epsilon>0$, metric-TSP interdiction admits a
deterministic $(4+\epsilon)$-approximation respecting the original
budget, with running time $N^{O(1+1/\epsilon)}$.
\end{corollary}

\begin{proof}
Handle an affordable disconnection by a minimum cut. Otherwise apply
Theorem~\ref{thm:approximation} with parameter $\epsilon/2$, and let
$R$ be its output. If $R^\star$ is an optimal deletion for the
metric-TSP objective, then
\[
\TSP(G-R)\ge F(R)
\ge\frac{F(R^\star)}{2+\epsilon/2}
\ge\frac{\TSP(G-R^\star)}{4+\epsilon}.
\]
The algorithm compares candidates using MST values, so it does not
need to solve a TSP instance.
\end{proof}

More generally,~\eqref{eq:tsp-bounds} turns an
$\alpha$-approximation for MST interdiction into a
$2\alpha$-approximation for this metric-TSP objective.
These inequalities alone do not transfer the hardness result of
Theorem~\ref{thm:hardness} to metric-TSP interdiction.

\section{Conclusion}

\edit{
We presented two deterministic $(2+\epsilon)$-approximation algorithms for MST interdiction, improving the previous factor of $4$ while respecting the original deletion budget. The algorithms use different approaches: one combines two weight thresholds with Lagrangian rounding, while the other decomposes the problem into heavy connections and bounded-weight interdiction within vertex partitions. Together with our conditional inapproximability result under the Small Set Expansion Hypothesis, these algorithms establish a nearly tight approximation ratio. They also yield a $(4+\epsilon)$-approximation for metric-TSP interdiction.
}


An exact factor-$2$ approximation remains open. Another question is
whether the dependence on $1/\epsilon$ can be removed from the
exponent of $N$. \edit{For metric-TSP interdiction, it would be interesting to improve
the $(4+\epsilon)$ guarantee beyond the direct MST comparison.}

\section*{Acknowledgements}
The authors used ChatGPT to assist with literature exploration, the development and refinement of algorithmic ideas
and mathematical arguments, and improvements to the manuscript's
organization and language. The authors take full
responsibility for the final manuscript, including the correctness of the results and the accuracy of the references.


\appendix
\section{Affordable-Partition Enumeration: Proof of Theorem~\ref{thm:partitions}}
\label{app:packing}

We prove Theorem~\ref{thm:partitions}, making explicit the handling of
zero costs, disconnected graphs, and exact rational computation. The
argument uses forest packing in the same general spirit as the
tree-packing approach to cut enumeration~\cite{CQX20}.
We first handle zero budget, then establish and compute a sufficiently
large forest packing. Finally, we recover each affordable partition
by removing a few edges from one support forest and grouping the
resulting components.

\begin{proof}[Proof of Theorem~\ref{thm:partitions}]
Write $H=(V,F)$ and $m=|F|$, after ignoring loops. An affordable
partition cannot have more than $k$ parts: otherwise, grouping its
parts into $k+1$ nonempty parts would give a $(k+1)$-cut of cost at
most $C$.

\paragraph{The case $C=0$.}
Let $H^+=(V,\{e\in F:c_e>0\})$. Its number $r$ of connected
components is at most $k$. Indeed, if $r\ge k+1$, grouping these
components into $k+1$ parts would produce a zero-cost $(k+1)$-cut.
A partition has crossing cost zero if and only if every component of
$H^+$ is contained in one of its parts. Thus all affordable partitions
are exactly the groupings of these $r$ components. Enumerating their
at most $r^r\le k^k$ groupings proves the theorem in this case.

Assume henceforth that $C>0$.

\paragraph{A packing of spanning forests.}
The hypothesis also implies that $H$ itself has at most $k$ connected
components, since otherwise a $(k+1)$-cut of cost zero would exist.
Hence the family $\mathcal F_k$ of spanning forests of $H$ with
exactly $k$ components is nonempty. Each such forest has $n-k\ge1$
edges. Consider the packing linear program
\begin{equation}
\label{eq:forest-packing}
\begin{aligned}
\nu=\max\quad &\sum_{T\in\mathcal F_k}y_T\\
\text{subject to}\quad
&\sum_{T\in\mathcal F_k:e\in T}y_T\le c_e &&(e\in F),\\
&y_T\ge0 &&(T\in\mathcal F_k).
\end{aligned}
\end{equation}
Its dual is
\begin{equation}
\label{eq:forest-packing-dual}
\begin{aligned}
\min\quad &\sum_{e\in F}c_e\ell_e\\
\text{subject to}\quad
&\sum_{e\in T}\ell_e\ge1 &&(T\in\mathcal F_k),\\
&\ell_e\ge0 &&(e\in F).
\end{aligned}
\end{equation}
The packing is feasible and bounded: summing its capacity constraints
gives $(n-k)\sum_Ty_T\le\sum_ec_e$. The dual is feasible, for example
with every $\ell_e=1$.

We claim that
\begin{equation}
\label{eq:forest-packing-value}
  \nu\ge\frac{C}{2k}.
\end{equation}
Take a feasible dual vector $\ell$. For $t\ge0$, define
\[
  H_t=(V,\{e\in F:\ell_e\le t\}),
  \qquad r(t)=\kappa(H_t).
\]
If $r(t)>k$, contract the components of $H_t$, and let $C_t$ be the
total cost of the edges between distinct contracted vertices. Each
such edge has $\ell_e>t$. The $k$ contracted vertices of smallest
weighted degree have total degree at most $2kC_t/r(t)$. Isolating
these vertices and grouping all the others into one part produces a
$(k+1)$-cut of $H$ with cost at most this amount. By the hypothesis,
\[
  C<\frac{2kC_t}{r(t)}.
\]
Consequently, for every $t\ge0$,
\begin{equation}
\label{eq:forest-layer-bound}
 c(\{e\in F:\ell_e>t\})
 \ge \frac{C}{2k}\bigl(r(t)-k\bigr)_+,
\end{equation}
where $(x)_+=\max\{x,0\}$. For $r(t)>k$, this follows from
$c(\{e:\ell_e>t\})\ge C_t>Cr(t)/(2k)$; for $r(t)\le k$, the
right-hand side is zero.

Integrating~\eqref{eq:forest-layer-bound} gives
\begin{equation}
\label{eq:forest-integral}
\begin{aligned}
 \sum_{e\in F}c_e\ell_e
 &=\int_0^\infty c(\{e\in F:\ell_e>t\})\,dt\\
 &\ge\frac{C}{2k}\int_0^\infty\bigl(r(t)-k\bigr)_+\,dt
 \ge\frac{C}{2k}.
\end{aligned}
\end{equation}
To justify the last inequality, Kruskal's algorithm stopped when
$k$ components remain produces a minimum-$\ell$ spanning forest with
$k$ components. The integral in~\eqref{eq:forest-integral} is exactly
the weight of this forest: it sums the selected edge lengths by
integrating the number of selected edges whose lengths exceed $t$.
This identity remains valid for disconnected $H$, because $H$ has
at most $k$ components. Dual feasibility makes the minimum forest
weight at least one. Strong linear-programming duality now
proves~\eqref{eq:forest-packing-value}.

\paragraph{Computing a packing with polynomially many forests.}
The separation problem for~\eqref{eq:forest-packing-dual} is a
minimum-$\ell$ spanning forest with $k$ components, which the same
stopped Kruskal algorithm computes in polynomial time. All numerical
data are rational. Moreover, the dual may be restricted to
$0\le\ell_e\le1$: capping a coordinate at one preserves every
constraint $\sum_{e\in T}\ell_e\ge1$ and cannot increase the objective.
The optimization--separation equivalence, with primal-solution
recovery, therefore computes an exact optimal packing represented
by polynomially many forests~\cite{GLS81}. More explicitly, recover the forests defining the required constraints
of the bounded dual, and solve the original packing LP using only
these forests as variables. Capping at one also preserves every restricted
forest constraint, so the restricted packing retains the same optimum.
Choosing a basic optimum
over the recovered columns leaves at most $m$ positive forest
variables, since the packing has $m$ capacity constraints. Its
encoding length is polynomial in $N$, as its constraint matrix has
$0/1$ entries and its capacities are rational input data. Write
$\mathcal S\subseteq\mathcal F_k$ for this support. These arguments
allow zero-cost edges and require no connectivity assumption beyond
$\kappa(H)\le k$.

\paragraph{Recovering every affordable partition.}
Fix an affordable partition $\mathcal P$, and put
$D=\delta_H(\mathcal P)$. The packing constraints give
\[
 \sum_{T\in\mathcal S}y_T|T\cap D|
 =\sum_{e\in D}\sum_{T\in\mathcal S:e\in T}y_T
 \le c(D)\le C.
\]
Since $\sum_{T\in\mathcal S}y_T=\nu\ge C/(2k)>0$, some
$T\in\mathcal S$ satisfies $|T\cap D|\le2k$. Deleting $T\cap D$
from $T$ leaves at most $3k$ components, each contained in a part of
$\mathcal P$. Thus grouping these forest components recovers
$\mathcal P$.

For every $T\in\mathcal S$, enumerate all subsets $Z\subseteq T$
with $|Z|\le2k$. For each, enumerate all groupings of the components
of $T-Z$ into at most $k$ nonempty parts, and retain exactly the
partitions of crossing cost at most $C$. The preceding argument
proves completeness. Canonically ordering the parts permits removal
of duplicate partitions.

A support forest has $n-k$ edges, so the edge-subset enumeration
has $n^{O(k)}$ choices. There are at most $3k$ components to group,
giving at most $k^{O(k)}$ choices. Because $k<n$, this factor is
absorbed into $n^{O(k)}$. The support size, component computations,
cost comparisons, and exact arithmetic contribute only polynomial
factors in $N$. This proves the stated deterministic running time.
\end{proof}

\section{\edit{Technical Tools for the Residual Rounding Bound}}
\label{app:rounding-tools}

\edit{This appendix provides the technical tools used in the proof of
Lemma~\ref{lem:baseline}. Section~\ref{app:relaxation} computes the
Lagrangian bound and the nested deletion sets, and
Section~\ref{app:tree-rounding} proves the tree-knapsack rounding lemma.}

\subsection{\edit{Computing the Lagrangian Bound and Nested Deletion Sets}}
\label{app:relaxation}

This subsection supplies the optimization step used in
Lemma~\ref{lem:baseline}. We compute $L(D)$ exactly and either obtain
a feasible set of value $L(D)$ or two nested sets whose costs lie on
opposite sides of $D$. The combinatorial rounding of these sets is
given in Section~\ref{subsec:rounding-proof}.

Use the notation of Lemma~\ref{lem:baseline}, and write $n=|V(H)|$.
The cases $t=0$, $n\le1$, and $J=\varnothing$ require no optimization:
the empty set attains $L(D)=h(\varnothing)$. Assume otherwise.

\paragraph{A value oracle and a separation oracle.}
The threshold identity expresses $h(S)$ as a nonnegative weighted
sum of component counts. Each such count is nondecreasing and
supermodular in the deleted set. For completeness, if $F_s$ is the
edge set of $H_{<s}$ and $r$ is the graphic matroid rank, then
\[
\kappa(H_{<s}-S)=n-r(F_s\setminus S).
\]
Submodularity of $r$ gives supermodularity of this component count.
Thus $h$ is nondecreasing and supermodular.

For fixed rational $\lambda\ge0$, maximizing
$h(S)-\lambda c(S)$ is equivalent to minimizing the submodular
function $\lambda c(S)-h(S)$. An MST computation evaluates $h(S)$,
so exact submodular function minimization gives the maximizing value
and a maximizing set in polynomial time~\cite{IFF01,Sch00}.
It also separates the constraints in
\[
\min\ \lambda D+z
\quad\text{subject to}\quad
\lambda\ge0,\qquad
z\ge h(S)-\lambda c(S)\quad(S\subseteq J).
\]

\paragraph{Bounding the variables and computing the optimum.}
Let $c_{\min}=\min_{e\in J}c_e>0$ and set
\[
\Lambda=(n-1)t/c_{\min}+1.
\]
For $\lambda\ge\Lambda$, every nonempty set has penalized value
less than zero, whereas the empty set has value
$h(\varnothing)\ge0$. For these multipliers the objective is
$\lambda D+h(\varnothing)$, which does not decrease with $\lambda$.
An optimum therefore exists with $0\le\lambda\le\Lambda$.
We may also impose $0\le z\le(n-1)t$.

All these bounds have polynomial binary encoding length for rational
input data. Exact separation and the optimization--separation
equivalence~\cite{GLS81} compute an optimal multiplier $\lambda^*$
and the value $L(D)$ in polynomial time.
When $D=0$, the preceding large-multiplier argument gives
$L(0)=h(\varnothing)$; return the empty deletion.

\paragraph{Recovering two nested maximizing sets.}
Suppose $D>0$. The sets maximizing
$h(S)-\lambda^*c(S)$ are closed under union and intersection.
Indeed, supermodularity applied to two maximizers implies that their
union and intersection are also maximizers.
Let $S_-$ be the intersection of all maximizers and $S_+$ their
union. Then both are maximizers and $S_-\subseteq S_+$.

They can be recovered with at most $2|J|$ additional optimization
calls. An edge lies in $S_-$ exactly when forbidding it decreases
the maximizing value. It lies in $S_+$ exactly when requiring it
preserves that value. Each requirement or prohibition gives a
submodular minimization problem on a smaller ground set.

\paragraph{Locating the budget between the two costs.}
In~\eqref{eq:round-lagrangian}, each set $S$ defines an affine
function of $\lambda$ with slope $D-c(S)$.
Among the functions attaining the maximum at $\lambda^*$, the
largest slope is $D-c(S_-)$ and the smallest is $D-c(S_+)$.
If $\lambda^*=0$, the right derivative is nonnegative at the
minimum, so $c(S_-)\le D$. Also $h(S_-)=L(D)$, and we return $S_-$.

If $\lambda^*>0$, the left derivative is nonpositive and the right
derivative is nonnegative. Hence
\[
c(S_-)\le D\le c(S_+).
\]
If either equality holds, the corresponding set is feasible and has
value $L(D)$. Otherwise the costs strictly bracket $D$.
With
$\theta=(D-c(S_-))/(c(S_+)-c(S_-))$,
their common penalized value gives
\[
L(D)=(1-\theta)h(S_-)+\theta h(S_+),
\]
as used in~\eqref{eq:round-interpolation}.
The rational linear programs and oracle calls above have polynomial
encoding length, so the entire procedure is polynomial-time.

\subsection{\edit{Tree-Knapsack Rounding}}
\label{app:tree-rounding}

For completeness, we give the tree-knapsack rounding argument used in
Lemma~\ref{lem:baseline}. It follows the rounding framework of
Linhares and Swamy~\cite{LS17}. At each stage, all unresolved
fractional choices lie in one branch. We discard its top node and
reoptimize below it. The discarded nodes lie on one root-to-leaf
path, which bounds the total profit lost.

\begin{lemma}[Tree-knapsack rounding]
\label{lem:tree}
Consider a rooted tree with nonnegative rational profits $p_v$ and costs
$d_v$ on its nonroot nodes. A selection must contain every descendant of
each selected node. Suppose that the total profit on every root-to-leaf
path is at most $t$. For any rational budget $D\ge0$, an integral
feasible selection whose profit is at least the optimum of
\begin{equation}
\label{eq:tree-relaxation}
\begin{aligned}
\max\quad &\sum_v p_vx_v\\
\text{subject to}\quad
&\sum_v d_vx_v\le D,\qquad 0\le x_v\le1,\\
&x_v\le x_u
\quad\text{whenever $u$ is a child of a nonroot node $v$}
\end{aligned}
\end{equation}
minus $t$ can be found in polynomial time. All sums exclude the root.
\end{lemma}

\begin{proof}
Choose an extreme optimal solution $x$ of~\eqref{eq:tree-relaxation}.
If it is integral, return it.

\paragraph{Fractional choices lie in one branch.}
We first show that at most one subtree
rooted at a child of the root contains fractional coordinates.

Suppose that two such subtrees contain fractional coordinates. In each,
choose a maximal connected set of nodes with a common fractional value;
call these sets $A_1$ and $A_2$. Changing all coordinates in one $A_j$
by the same sufficiently small amount, in either direction, preserves
the parent--child inequalities within that set. Every inequality joining
it to another node has positive slack, since adjacent nodes with the same
value would belong to the same connected set. A sufficiently small
change also preserves these inequalities and the bounds $0<x_v<1$.

Let $d(A_j)=\sum_{v\in A_j}d_v$. If one of these costs is zero, increase
and decrease the coordinates in that set by equal small amounts. This
produces two distinct feasible points whose average is $x$, contradicting
extremality. If both costs are positive, increase the coordinates in
$A_1$ and decrease those in $A_2$, choosing the changes so that their
total cost effects cancel. Reversing both changes again gives two
distinct feasible points whose average is $x$. This is also a
contradiction. Hence only one root-child subtree can contain fractional
coordinates.

\paragraph{One rounding step.}
Let $v$ be that child. Fix all coordinates outside its subtree at their
current integral values, and set $x_v=0$. Reoptimize over the proper
descendants of $v$, using the original budget minus the cost of the fixed
selection outside the subtree. Regard these descendants as a forest
attached to a new dummy root, and choose an extreme optimum of the new
linear program. Setting $x_v=0$ saves cost and relaxes its inequalities
with its children, so the old descendant coordinates remain feasible.
The reoptimization therefore loses no further profit. The profit lost
in this step is at most $p_vx_v\le p_v$, where $x_v$ denotes the value
before it was set to zero.

\paragraph{Bounding the total loss.}
Apply the same procedure to the new extreme optimum. At each step,
integral coordinates outside the unique subtree containing fractional
coordinates are fixed, its root is set to zero, and its proper
descendants are reoptimized. If a reoptimization is integral, the
procedure stops. Otherwise the active subtree strictly shrinks.
Consequently, the nodes set to zero lie on one root-to-leaf path of the
original tree, and the sum of the charged losses is at most $t$.

\paragraph{Feasibility and running time.}
Every reoptimization uses the budget remaining after the fixed
selection, so the final selection respects the budget. The
parent--child inequalities remain valid: those inside a fixed subtree
are unchanged, those inside the active subtree are imposed by its linear
program, and the boundary parent has been set to zero. Thus the final
integral selection contains every descendant of each selected node.
There are at most the depth of the original tree many reoptimizations,
each a linear program of polynomial size with rational data. The
procedure is therefore polynomial-time and loses at most $t$ in profit.
\end{proof}

\section{Proof of the Conditional Hardness Result}
\label{app:hardness}

We first check the budget endpoint in the earlier Maximum Components
proof. We then give a self-contained proof of
Theorem~\ref{thm:hardness} using a normalized expansion promise,
Manurangsi's small-parts counting argument, and the value-preserving
transformation~\eqref{eq:components-transfer}.

\subsection{The budget boundary in the earlier proof}
\label{app:hardness-comparison}

In the soundness proof of~\cite[Theorem~5, p.~21]{AACR22}, the
displayed estimate gives cut cost at least $B$, but the subsequent
conclusion excludes cuts of cost at most $B$. That estimate does not
exclude equality. This is a repairable strictness omission, not a
counterexample to the theorem. We show that the same counting argument
gives the strict inequality needed for an inclusive budget.

Use a $d$-regular source graph $K$ on $n$ vertices, where $d>0$.
Put $s=\delta n$, with $0<\delta\le1/2$ and $s$ integral, and fix
$0<\rho<1$ and $0<\eta<\rho/4$. Consider a partition of the vertex set of $K$ into $k>(1/2+\rho/2)s$ nonempty parts, ordered by increasing size.
Let $A=T_1\cup\cdots\cup T_i$ be the longest initial union of size
at most $s$, allowing $i=0$. There is a next part since $s<n$.
The stronger assertion needed for an inclusive budget is
\[
|A|>(1/2+2\eta)s.
\]
To prove it, suppose instead that $|A|\le(1/2+2\eta)s$.
Then $i\le|A|\le(1/2+2\eta)s$, while every remaining part has size
at least $|T_{i+1}|>s-|A|\ge(1/2-2\eta)s$. Consequently,
\[
n\ge(k-i)|T_{i+1}|
>(\rho/2-2\eta)(1/2-2\eta)\delta^2 n^2,
\]
which is impossible for
$n>[(\rho/2-2\eta)(1/2-2\eta)\delta^2]^{-1}$.
Both factors are positive. The asserted strict lower bound on
$|A|$ follows.

In a NO instance, extend $A$ to a set of exactly $s$ vertices.
The larger set has at least $(1-\eta)ds$ boundary edges, and adding
$s-|A|$ vertices changes the boundary by at most $d(s-|A|)$.
It follows that
\[
|\delta_K(A)|\ge d(|A|-\eta s)
>(1/2+\eta)ds=B.
\]
All these edges cross the partition, so its cost exceeds $B$.
The crucial change is to include equality in the contradictory
assumption on $|A|$. Thus the proof can be repaired without changing
the claimed hardness threshold. In the proof below, we instead use
Lemma~\ref{lem:hardness-small-parts} to obtain an explicit component
bound and handle the size conditions directly.

\subsection{The source problem and a counting lemma}

We use the following normalized formulation of the Small Set Expansion
Hypothesis (SSEH)~\cite{RS10,RST12}. For every rational
$0<\eta<1/2$, there is a constant $0<\delta\le 1/2$ such that the
following promise problem is NP-hard. The input is a symmetric matrix
$A=(a_{uv})_{u,v\in U}$ of nonnegative rational numbers, encoded in
binary, with
\[
\sum_{v\in U}a_{uv}=1\qquad\text{for every }u\in U.
\]
Write $n=|U|$, and assume that $s=\delta n$ is an integer. For
$X\subseteq U$, define
\[
\partial_A(X):=\sum_{u\in X,\ v\notin X}a_{uv}.
\]
The task is to distinguish the following cases:
\begin{description}
\item[YES:] There is a set $X\subseteq U$ with $|X|=s$ and
  $\partial_A(X)\le\eta s$.
\item[NO:] Every set $X\subseteq U$ with $|X|=s$ satisfies
  $\partial_A(X)\ge(1-\eta)s$.
\end{description}
This formulation follows from the usual regular-graph formulation by
dividing the weighted adjacency matrix by its common weighted degree.
The normalized expansion of $X$ is then $\partial_A(X)/|X|$.
We allow diagonal entries, counted once in each row sum; this does not
strengthen the assumption, since the normalized instances from the
usual formulation are still included.

We first isolate the elementary counting observation underlying the
soundness argument. Its proof is the small-parts argument from
Manurangsi~\cite[Section~2]{Man18}, stated for an arbitrary number of
parts.

\begin{lemma}
\label{lem:hardness-small-parts}
If an $n$-element set is partitioned into $k\ge2$ nonempty parts, then
some union $X$ of parts satisfies
\[
k-1-\sqrt n\le |X|\le k-1.
\]
\end{lemma}

\begin{proof}
Order the parts by nondecreasing size, and let $X$ be the union of the longest initial sequence with total size at most $k-1$. There is a next part because $k-1<n$. Suppose that $|X|<k-1-\sqrt n$. By the definition of $X$, the next part, and hence every subsequent part, has size greater than $k-1-|X|>\sqrt n$. Since each part is nonempty, at most $|X|$ parts have been selected to form $X$. As a consequence, more than $\sqrt n+1$ parts remain. Together these contain more than $n$ elements, a contradiction. Therefore, $ k-1-\sqrt n
\le|X|\le k-1$. This completes the proof of the lemma.
\end{proof}

\subsection{The reduction and the approximation gap}

\begin{proof}[Proof of Theorem~\ref{thm:hardness}]
Fix $\epsilon\in(0,1)$ and a rational constant
$0<\eta\le\epsilon/40$. Choose $\delta$ from SSEH, and consider an
instance $A$ of the corresponding promise problem. We may assume
\begin{equation}
\label{eq:hardness-large-instance}
\sqrt n\le\eta s,\qquad 1\le\eta s.
\end{equation}
Indeed, both inequalities hold whenever
$n\ge\lceil(\eta\delta)^{-2}\rceil$. For smaller $n$, the source
promise problem can be decided by enumerating all $s$-element subsets.
Since the bound on $n$ is constant, this enumeration takes polynomial
time in the binary encoding length. Thus an approximation algorithm
for the constructed instances below would also decide every source
instance in polynomial time.

 Add a new element $r$ to $U$ to form the vertex set of the input graph $G$ of the MST interdiction instance. For each unordered pair of distinct vertices $u,v\in U$ with $a_{uv}>0$, add an edge of weight zero and
deletion cost $a_{uv}$. Add an edge $ru$ of weight one for every
$u\in U$. Set
\[
B=\left(\frac12+\eta\right)s,
\]
and give each star edge $ru$ deletion cost $B+1$. No feasible deletion
contains a star edge, so every feasible residual graph is connected.
The constructed graph is simple, and all its deletion costs are
positive rational numbers of polynomial encoding length.

For a feasible deletion $R$, let $q(R)$ be the number of connected
components of the remaining zero-weight graph on $U$. Every spanning
tree must connect each of these components to $r$ using a weight-one
edge. Conversely, a zero-weight spanning tree inside each component,
together with one star edge per component, spans the constructed
graph. Consequently,
\begin{equation}
\label{eq:hardness-component-value}
\MST(G-R)=q(R).
\end{equation}

Suppose that $A$ is a YES instance, and let $X$ be a set of size $s$
with $\partial_A(X)\le\eta s$. Delete every zero-weight edge incident
to $X$. Summing the row sums over $X$ shows that the deletion cost is
\[
\frac{s+\partial_A(X)-\sum_{u\in X}a_{uu}}{2}
\le\frac{(1+\eta)s}{2}
\le B.
\]
All $s$ vertices of $X$ become isolated in the zero-weight graph.
Since $s\le n/2$, its complement is nonempty and contributes at least
one additional component. By~\eqref{eq:hardness-component-value},
\begin{equation}
\label{eq:hardness-yes}
\OPT\ge s+1.
\end{equation}

Suppose that $A$ is a NO instance. The promise also bounds the total
weight inside any set $X$ of size at most $s$: extend $X$ to a set
$Y$ of size $s$ and use nonnegativity to obtain
\[
\sum_{u,v\in X}a_{uv}
\le\sum_{u,v\in Y}a_{uv}
=s-\partial_A(Y)
\le\eta s.
\]
The row-sum normalization therefore gives
\begin{equation}
\label{eq:hardness-small-set-boundary}
\partial_A(X)
=|X|-\sum_{u,v\in X}a_{uv}
\ge |X|-\eta s
\qquad\text{whenever }|X|\le s.
\end{equation}
If $X$ is a union of residual zero-weight components, all edges
crossing its boundary have been deleted. Hence
$c(R)\ge\partial_A(X)$ for every such union.

Fix a feasible deletion $R$, and write $q=q(R)$. First suppose that
$q\ge s+1$. Group its components into $s+1$ nonempty parts.
Lemma~\ref{lem:hardness-small-parts} supplies a union $X$ of these
parts with $s-\sqrt n\le|X|\le s$. Equations
\eqref{eq:hardness-small-set-boundary}
and~\eqref{eq:hardness-large-instance} give
\[
c(R)\ge |X|-\eta s
\ge(1-2\eta)s
>\left(\frac12+\eta\right)s=B,
\]
where the strict inequality follows from $\eta<1/6$. This contradicts
feasibility, so $q\le s$.

If $q\ge2$, apply Lemma~\ref{lem:hardness-small-parts} directly to
the $q$ components. The resulting union $X$ satisfies
$q-1-\sqrt n\le|X|\le q-1\le s$. Thus
\[
B\ge c(R)\ge |X|-\eta s
\ge q-1-\sqrt n-\eta s.
\]
Using~\eqref{eq:hardness-large-instance} once more yields
\begin{equation}
\label{eq:hardness-no}
q\le\left(\frac12+2\eta\right)s+\sqrt n+1
\le\left(\frac12+4\eta\right)s.
\end{equation}
This bound also holds for $q=1$, since $1\le\eta s$.
By~\eqref{eq:hardness-component-value}, it bounds the value of every
feasible interdiction in the NO case.

Finally,
\[
(2-\epsilon)\left(\frac12+4\eta\right)
\le(2-\epsilon)\left(\frac12+\frac{\epsilon}{10}\right)
<1.
\]
A $(2-\epsilon)$-approximation must therefore return a value greater
than $(1/2+4\eta)s$ in the YES case, by~\eqref{eq:hardness-yes}.
In the NO case, no feasible deletion exceeds this threshold. The
construction uses $n+1$ vertices and at most $\binom n2+n$ edges, and
the returned value can be checked by an MST computation. An assumed
deterministic polynomial-time approximation would thus decide the
NP-hard source promise problem in polynomial time, implying
$\mathsf{P}=\mathsf{NP}$.
\end{proof}

\end{document}